\documentclass[twocolumn]{emulateapj}
\usepackage{natbib}

\shorttitle{The suppression of fragmentation by dark matter annihilation}
\shortauthors{Smith et al.}

\def\lsim{\mathrel{\raise.3ex\hbox{$<$\kern-.75em\lower1ex\hbox{$\sim$}}}}
\def\gsim{\mathrel{\raise.3ex\hbox{$>$\kern-.75em\lower1ex\hbox{$\sim$}}}}

\newcommand{\eq}{Equation }
\newcommand{\fig}{Figure }
\newcommand{\sect}{Section }
\newcommand{\tab}{Table }

\newcommand{\msun}{\,M$_{\odot}$ }

\newcommand{\myr}{\,M$_{\odot}$\,yr$^{-1}$}

\newcommand{\cmc}{\,cm$^{-3}$}
\newcommand{\kms}{\,kms$^{-1}$}

\newcommand{\E}{\times 10}

\begin{document}
% This is version 2.1 as it includes Fabios changes.

\title{WIMP DM and first stars: suppression of fragmentation in primordial star formation}

\author{Rowan J. Smith$^{1}$, Fabio Iocco$^{2}$, Simon C. O. Glover$^{1}$, Dominik R. G. Schleicher$^{3}$, Ralf S. Klessen$^{1}$, Shingo Hirano$^{4}$, Naoki Yoshida$^{4}$}
\affil{1. Universit\"at Heidelberg, Zentrum f\"ur Astronomie, Institut f\"ur Theoretische Astrophysik, Albert-Ueberle-Str. 2, 69120 Heidelberg, Germany 2. The Oskar Klein Centre, Department of Physics, AlbaNova, Stockholm University, SE-106 91 Stockholm, Sweden 3. Institut f\"ur Astrophysik, Georg-August-Universit\"at, Friedrich-Hund-Platz 1, 37077 G\"ottingen, Germany 4. Institute for the Physics and Mathematics of the Universe, University of Tokyo, 5-1-5 Kashiwanoha, Kashiwa, Chiba 277-8568, Japan} 
% ***some of these need converting to full versions***

\email{rowan@uni-heidelberg.de}

\begin{abstract}
We present the first 3D simulations to include the effects of dark matter annihilation feedback during the collapse of primordial mini-halos. We begin our simulations from cosmological initial conditions and account for  dark matter annihilation in our treatment of the chemical and thermal evolution of the gas. The dark matter is modelled using an analytical density profile that responds to changes in the peak gas density. We find that the gas can collapse to high densities despite the additional energy input from the dark matter. No objects supported purely by dark matter annihilation heating are formed in our simulations. However, we find that the dark matter annihilation heating has a large effect on the evolution of the gas following the formation of the first protostar. Previous simulations without dark matter annihilation found that protostellar discs around Population III stars rapidly fragmented, forming multiple protostars that underwent mergers or ejections. When dark matter annihilation is included, however, these discs become stable to radii of 1000~AU or more. In the cases where fragmentation does occur, it is a wide binary that is formed. %Our work does not include a live dark matter halo and it is possible that interactions between the baryons and dark matter could decrease the latter's central density. However, the timescales for this process are sufficiently long that the central star would be nearing the stage where ionisation feedback becomes important when this occurs.
\end{abstract}


\section{Introduction}\label{intro}
Modern cosmology describes a Universe in accelerated expansion, with 73\% of its energy density today consisting of an unknown substance dubbed ``dark energy'', and the remaining 27\% consisting primarily of matter, with a very minor contribution from radiation. It is remarkable that of the whole matter content, only 17\% of it can be accounted for by known particles described by the standard model of particle physics, whereas the remainder is of unknown nature, only (at most) weakly interacting with its surroundings. The nature of this dark matter (DM) is still unknown, and theories such as supersymmetry that extend the standard model and provide candidates for the particle making up the DM are currently being tested at the Large Hadron Collider at CERN.

In spite of the unknown nature of the basic constituents of our Universe, its description within the standard cosmological model holds solidly. The growth of structure is predicted to take place in a hierarchical fashion. Smaller DM and gas structures (halos) are predicted to collapse gravitationally at earlier times than more massive ones, and evidence for this is indeed observed. 

So far, however, the first generation of stars to have formed in the Universe, after the long period of cosmic silence and darkness known as the cosmic ``dark ages'' has not been observed. The formation of stars from this first generation -- Population III stars -- is harboured in the bosom of the very first DM and gas halos to undergo gravitational collapse: structures of masses 10$^5 \lsim $~M/M$_\odot\lsim$~10$^7$ which become nonlinear at redshifts $30 \lsim z \lsim 20$. These stars are too distant to be observed directly with any current telescope, and so their nature must be probed theoretically. Originally, it was thought that isolated, extremely massive stars \citep{Abel00,Abel02,Yoshida08} would be formed from such primordial halos as they have high temperatures and only simple molecules such as H$_2$ and HD to provide the necessary cooling \citep[e.g.][]{Glover05,Omukai05}. However, recent work has shown even in these conditions small stellar systems may be formed due to fragmentation occurring in protostellar discs around Population III stars \citep{Clark08,Turk09,Stacy10,Clark11a,Clark11b,Greif11,Greif12}. Such systems are predicted to have masses ranging from a few tenths to a few tens of a solar mass, with a mass spectrum that is presumably flat \citep{Dopcke12}.

Interestingly enough, the story of the first stars could be intimately related to that of the DM.  One popular class of candidates for the DM particle are the weakly interacting massive particles (WIMPs). These are the lightest supersymmetric partners in supersymmetry theories conserving $R$-parity, or the lightest mode in extra dimension theories under $T$-parity conservation. They are stable and they ``naturally'' comply with all the phenomenological requirements of DM \citep{Taoso07}. Although we refer the reader to \citet{Bertone04} for a detailed review of WIMPs, we note here that they have two very important properties: they are self-annihilating Majorana particles (i.e.\ they are their own anti-particles) and they interact only weakly with baryons.

While the effects of WIMPs on stellar formation in the local Universe are suppressed \citep{Ascasibar06}, \citet{Spolyar08} noted that the annihilation of WIMPs accumulated by gas-driven gravitational drag into the center of a collapsing halo would strongly heat the gas at the center of the halo. In the absence of metals -- i.e.\ in a halo forming Population III stars -- the energy input could potentially be so large as to overcome the cooling provided by H$_{2}$ molecules. The authors speculated that this could halt the collapse of the gas, and form what they dubbed a ``dark star'': a pseudo-star with a size of an AU or more powered by DMA rather than by nuclear fusion. It was also soon realised that if a canonical protostar were to be formed, it could capture DM through weak scattering between the WIMPs and the baryons constituting the star, and that this could in principle affect the main sequence (MS) evolution of the newly formed star \citep{Iocco08a}.
 
 Several studies have since addressed these processes in more detail. \citet{Freese08} and \citet{Spolyar09} proposed that the accretion of infalling gas could be sustained by DMA indefinitely up to the formation of objects of masses of 10$^5$M$_\odot$. However, \citet{Iocco08b} found that such a phase could not last for longer than 10$^4$ yr before reaching the Hayashi track, and the findings in \citet{Hirano11} were intermediate between these extremes. The results of those studying the capture of DM onto a hydrostatic object mediated via weak interaction, are more homogenous. It is found that weakly captured DM can in principle indefinitely power the star as it is sustained not by nuclear reactions but by DMA, leading to a prolongation of the star's life \citep{Iocco08b, Yoon08, Taoso08}. It is not clear, however, how much DM can be consumed in the pre-MS process, and therefore how much is available to be captured during the main sequence \citep{Sivertsson10}. Indeed, some of the more extreme models of dark star formation create problems in standard reionisation scenarios due to their predictions of very high stellar masses or very long lifetimes \citep{Schleicher08,Schleicher09}. However, in some cases dark stars may delay rather than accelerate the reionisation process \citep{Scott11}. Direct observation seems challenging for most models of dark stars, \cite{Zackrisson10b}, whereas the most extreme objects can already be ruled out with the use of HST observations, \citep{Zackrisson10a}.

Regardless of the efficiencies of these processes, DMA takes place throughout the halo. The effects of DMA will be felt most keenly when the halo has collapsed to high central gas densities, but subtle effects due to the influence of annihilation upon the gas chemistry can take place even earlier in the halo's life. This has been addressed in \citet{Ripamonti10}, who used a one-dimensional code to self-consistently solve for the chemical, thermal and dynamical evolution of the cloud in the presence of DMA. Intriguingly, the authors found that DMA had little effect on the gas prior to its collapse to near-stellar densities. However, the scope of these results was limited as the maximum resolvable density in the \citet{Ripamonti10} study was reached {\it before} a hydrostatic object was formed. In complementary work, \citet{Stacy12} addressed the phases {\it after} the formation of a hydrostatic object, adopting an ad-hoc prescription for the earlier phase. These authors found that if multiple protostars were present, dynamical interactions would displace them from the DM density peak, thereby removing them from the region where DMA would have the most influence.

In this paper we bridge the gap before and after star formation in one continuous three-dimensional simulation that fully accounts for the effects of DMA, primordial gas chemistry and time-dependent heating and cooling. This allows us to determine at which stage, if any, DMA plays a role in the formation of a Population III star. Our paper is structured as follows: \sect \ref{method} outlines the method, \sect \ref{ics} describes the initial conditions, Sections \ref{results1} and \ref{results2} present our results, Section \ref{discussion} discusses the results, and finally \sect \ref{conclusion} summarises our conclusions.

\section{Method}\label{method}
\subsection{Basic approach}
\label{basic}
We perform the calculations for this paper using the smoothed particle hydrodynamics (SPH) code {\sc gadget 2} \citep{Springel05}. We have substantially modified this code to include a fully time-dependent chemical network, details of which can be found in the Appendix of \citet[][]{Clark11a}. Our treatment includes: H$_2$ line cooling \citep{Glover08}, which is treated in the optically thick regime using the Sobolev approximation \citep{Yoshida06,Clark11a}; collision-induced emission from very high density H$_{2}$ \citep{Ripamonti04}; cooling due to the collisional ionisation and recombination of hydrogen and helium \citep{Glover07a}; as well as heating and cooling arising due to changes in the molecular fraction, shock heating, compressional heating and cooling due to rarefactions. \citet{Turk11} showed that the choice of the H$_2$ three-body formation rate coefficient influences the resulting dynamics of the gas within the halo. In this work we use the three-body H$_2$ formation rate of \citet{Glover08b} which is intermediate within the range of the published rates. 

To treat gas which collapses to scales smaller than the resolution limit of our simulations, we use a sink particle approach. We create sinks using the standard \citet{Bate95} algorithm. Our particular implementation of this is the same as that used in \citet{Jappsen05}. The minimum number density for sink particle creation is $10^{16} \:{\rm cm^{-3}}$, but it is important to note that candidate regions must also satisfy a series of tests to ensure that they are unambiguously collapsing. The ratio of thermal energy and rotational energy to the magnitude of the gravitational potential energy ($\alpha$ and $\beta$ respectively) for the particles that will be turned into the sink must satisfy the conditions
\begin{equation}
\alpha \leq \frac{1}{2},
\end{equation}
\begin{equation}
\alpha + \beta \leq 1.
\end{equation}

In addition, the total energy of the particles must be negative, and the divergence of their accelerations must be zero in order to prevent structures that are likely to bounce or be tidally disrupted being turned into sinks. Consequently sink particle formation often occurs above this density threshold.

We set the outer accretion radius of the sink particles, $r_{\rm out}$, to 6~AU and the inner accretion radius, $r_{\rm in}$, to 4~AU. SPH particles that come within a distance $r_{\rm out} > r > r_{\rm in}$ of a sink are accreted only if they are gravitationally bound to that sink. SPH particles that come closer than $r_{\rm in}$ are always accreted. Once sinks have formed, we account for the accretion luminosity produced by the ongoing accretion of gas onto the sinks using the scheme described in \citet{Smith11b}, which assumes that all of the protostars that form are normal Population III protostars, and not dark stars. We would expect the heating produced by accretion onto dark stars to be smaller, owing to their larger radii, and so this procedure gives us an upper limit on the effect of the accretion luminosity.

\subsection{Treatment of dark matter annihilation}
Most current simulations have insufficient resolution to follow the DM contraction within a primordial halo with the same resolution as that used for the baryons. For example, \citet{Abel02} had a DM mass resolution of 1.1 \msun which meant they could not determine the DM profile at radii smaller than around 0.05~pc. Instead, the method of adiabatic contraction developed by \citet{Blumenthal86} has been used by various authors \citep[e.g.][]{Spolyar08,Ripamonti10} to calculate how the DM distribution will respond to changes in the gravitational potential of the gas. \citet{Spolyar08} applied this method to the collapsing halo simulations of \citet{Abel02} and \citet{Gao07} and found that the DM density at the outer edge of the baryonic core after adiabatic contraction was well-fit by the expression
\begin{equation}\label{dmpeak}
\rho_{xc} \approx 5 n_{\rm p}^{0.81} \: {\rm GeV \: cm^{-3}},
\end{equation} 
where $\rho_{xc}$ is the DM density at the edge of the core and $n_p$ is the number density of hydrogen nuclei in the core.\footnote{To convert from the DM density in energy units to the value in the more familiar mass density units, one simply multiplies  by a factor of $c^{2}$. Moreover, it is also possible to express the gas density in energy units; for a gas with primordial composition, we have $\rho_{g} \simeq 1.24 \, n \: {\rm GeV} \: {\rm cm^{-3}}$, where $n$ is the number density of hydrogen nuclei.}

Given the considerable technical challenges involved if one attempts to follow the evolution of the DM density profile directly, we do not attempt to do so. Instead, we parameterise its effects, with the help of the adiabatic contraction results of \citet{Spolyar08}. The DM density profile is calculated within the code as follows. At each time-step, the point of maximum gravitational potential energy in the halo is found, which we assume to be the point at which the DM density has its maximum value. Next, the maximum baryon density of the halo is found and the DM density at the outer edge of the core calculated from it using \eq \ref{dmpeak}. Its radial dependence is then described  by a two-part power law fit, using power-law slopes drawn from the calculations of \citet{Ripamonti10}. In the outer halo, at distances $r > r_{c}$ from the centre of the density profile, we have
\begin{equation}\label{dmprofile}
\rho_{x} = \rho_0 \left(\frac{r}{1 \mathrm{pc}} \right)^{-1.8}  \: {\rm GeV \: cm^{-3}},
\end{equation} 
where $r$ is the radial distance from the centre and $\rho_{0} = 5 \times 10^{4} \: {\rm GeV \: cm^{-3}}$ is a normalising constant that is equal to the DM density at $r = 1 \: {\rm pc}$. The size of the central DM core, $r_{c}$, is determined by finding the radius at which the DM densities given by Equations~\ref{dmpeak} and \ref{dmprofile} are equal, which yields
\begin{equation}
r_{c} = 16.7 \left(\frac{n_{\rm p}}{10^{14} \: {\rm cm^{-3}}} \right)^{-0.81/1.8} \: {\rm AU}.
\end{equation}

Finally, the DM density profile within the core, at radii $r < r_{c}$, is given by
\begin{equation}\label{dmcore}
\rho_{x} = \rho_{xc} \left(\frac{r}{r_c} \right)^{-0.5}.
\end{equation} 
It is unclear how efficient adiabatic contraction will be within the central regions of the DM halo, given that gravitational collapse of the gas in these regions occurs rapidly. Our adoption of a peaked DM profile within the core may therefore be an overestimate; see, for instance, the discussion of the validity of the adiabatic contraction approximation given in \citet{Ripamonti10}. 

Given the DM density profile, the contribution of DMA to the halo energy budget is computed as follows. We follow \citet{Spolyar08} and adopt the standard DMA cross section of  $\langle\sigma v\rangle = 3\E^{-26}$ cm$^3$s$^{-1}$. For the DM particle mass, $m_{x}$, we note that astrophysical constraints require that $m_{x} \ge 10 \: {\rm GeV}$  \citep[see e.g.][]{Ackermann11,Galli11}. Additionally the desire to avoid fine-tuning while still explaining the gauge hierarchy problem argues for an upper limit $m_{x} \sim \mbox{a few} \: {\rm TeV}$. We therefore adopt a DM particle mass of $m_x=100$ GeV for our fiducial case, close to the middle of this range of values, but we also explore the effect of varying $m_x$. Following \citet{Valdes08}, we assume that one third of the energy from DMA is immediately carried away by neutrinos and that the remaining energy is split between direct heating of the gas, ionisation and dissociation of the constituents of the gas, and the excitation of H, He and H$_{2}$. The fraction that is deposited as heat is given by
\begin{equation}\label{fheat}
f_h=1-0.875(1-x_{\rm e}^{0.4052}),
\end{equation} 
and the fraction that goes into ionisation is given by
\begin{equation}\label{fion}
f_i=0.384(1-x_{\rm e}^{0.542})^{1.1952},
\end{equation} 
where $x_e \equiv n_{\rm e} / n$ is the fractional abundance of electrons. 
The remaining energy is radiated away in electronic transitions from H, He and H$_{2}$ and plays no direct role in the evolution of the gas. The total power per unit volume injected by DMA is
\begin{equation}
W_{\rm ann}=2  m_x \dot{N}_{\rm ann},
\end{equation}
where $\dot{N}_{\rm ann}$ is the number of DMAs per unit volume per unit time. This is given by
\begin{equation}\label{Nann}
\dot{N}_{\rm ann}=\frac{1}{2} n_x^2 \langle\sigma v\rangle,
\end{equation}
where $n_x=\rho_x/m_x$ is the number density of DM particles.
The resulting energy injection rates per unit volume for heating ($Q_h$) and ionisation ($Q_i$) are
\begin{equation}\label{Qheat}
Q_h=2 f_a \dot{N}_{\rm ann} (1-e^{-\tau_x})f_h m_x,
\end{equation} 
\begin{equation}\label{Qion}
Q_i=2 f_a \dot{N}_{\rm ann} (1-e^{-\tau_x})f_i m_x,
\end{equation}
where $f_a=2/3$ is the fraction of the total energy that affects the gas (one third of the energy escapes in the form of neutrinos). 

We estimate the optical depth of the gas to the annihilation products, $\tau_x$, by assuming that the 
baryonic density profile is flat within the DM core and declines as a power-law 
$\rho \propto r^{-2}$ at larger radii. For a profile of this form, the column density of gas, measured
radially outwards from a point $r$, is given at $r \leq r_{c}$ by
\begin{equation}
\Sigma(r) = \rho(r) \left[(r_{c} - r) + \left(r_{c} - \frac{r_{c}^2}{r_{h}}\right) \right]
\end{equation}
and at $r > r_{c}$ by
\begin{equation}
\Sigma(r)  = \rho(r) \left(r - \frac{r^2}{r_{h}}\right),
\end{equation}
where $r_{\rm h}$ is the virial radius of the halo, which we take to mark the edge of the gas distribution.
If we follow \citet{Ripamonti10}  and adopt a constant gas opacity of $\kappa=0.01$ cm$^2$g$^{-1}$,
and also restrict our attention to scales small compared to $r_{h}$, then we can write the optical depth as
\begin{equation}
\tau_x \equiv \kappa \Sigma(r) = \left \{ \begin{array}{lr}
\kappa \rho(r) (2r_{c} - r) & r \leq r_{c} \\
\kappa \rho(r) r & r > r_{c}
\end{array} 
\right.
\end{equation}
In practice, this expression for $\tau_x$ is an underestimate, for a couple of reasons. First, it assumes 
that the optical depth in any direction from point $r$ is the same as the value that we measure along a
ray passing radially outwards, when in reality the optical depths in other directions will be somewhat
larger. Second, the baryonic density profiles that we recover in our simulations do not completely match
up with the profile we assumed to derive $\tau_x$. Our assumed profiles and the real profiles both
have power-law slopes $\rho \propto r^{-2}$ in their outer regions, but the real profiles do not have the 
completely flat core that we assume within $r_{c}$.  Our estimate of $\tau_x$ is therefore too small at 
distances $r \ll r_c$, meaning that our derived heating and ionisation rates may also be too small.
In practice, we do not expect this to be a major source of error, as in the regime where DMA becomes important, gas with $r \ll r_c$ will typical have $\tau_x \gg 1$, meaning that
the heating and ionisation rates are insensitive to the precise value of $\tau_x$.

To convert from $Q_{i}$, the total energy per unit time per unit volume that goes into ionisations, to the
ionisation rates of the individual chemical species present in the gas, we follow the procedure outlined in \citet{Ripamonti07a}. They identified seven main reactions that can be caused by DMA: the ionisation of H, He, He$^+$ and D, and the dissociation of H$_2$, HD and H$_2^+$. Between them, these seven reactions account for the bulk of the energy lost in the form of ionisations or dissociations, and we follow \citet{Ripamonti07a} and split the energy available for ionisation between these seven species according to their relative abundances.

\section{Initial Conditions}\label{ics}

\begin{table}
	\caption{Overview of simulations}
	\centering
		\begin{tabular}{l c c}
   	         \hline
	         \hline
	         Simulation & Annihilation & DM particle mass [GeV] \\
	         \hline
	         H1-ref & no & 0\\
		 H1-lm & yes & 10\\
		 H1 & yes & 100 \\
		 H1-hm & yes & 1000\\
		 \hline		 H2-ref & no & 0 \\
		H2 & yes & 100 \\
		\hline
		\end{tabular}
	\label{sims}
\end{table}

We take our initial conditions from the cosmological simulations and resimulations by \citet{Greif11}. These simulations made use of the novel moving mesh code {\sc arepo} \citep{Springel10} to fully resolve the formation of five minihalos from cosmological initial conditions. Cells were refined during the evolution to ensure that the Jeans length was always resolved by at least 128 mesh points, up until the point at which the maximum number density in the  collapsing gas reached $n = 10^{9} \: {\rm cm^{-3}}$. All of the halos modelled by \citet{Greif11} formed multiple protostars with a range of masses. 

For this work we cut out the central two parsecs of the \citet{Greif11} simulations and continue their evolution using our modified version of {\sc gadget 2} with DMA, implemented as discussed in the previous section. The halos were selected at the point where their central gas number density  reached $n = 10^6 \: {\rm cm^{-3}}$ for the first time. We selected this point to begin our resimulation because preliminary modelling using simplified initial conditions showed that this was the point at which indirect feedback from DMA-induced ionisation first becomes important.

Each mesh point in {\sc arepo} is converted to an SPH particle with the same properties (mass, momentum, etc.) as the original mesh cell associated with that mesh point. As the network for primordial chemistry that is implemented in {\sc arepo} was derived from the network that we use in {\sc gadget}, both codes evolve the same set of chemical species and it is straightforward to transfer the chemical abundances from one code to  the other. The formation of the first protostar occurs in the central region of the halos where the SPH particle masses are $10^{-4} \: {\rm M_{\odot}}$, giving us a mass resolution of around $10^{-2}$ \msun \citep{Bate97}. For a test of the effectiveness of using {\sc arepo} initial conditions for {\sc gadget 2} see \citet{Smith11b}.

\citet{Greif11} simulated the evolution of five different DM halos. In this work, however, we focus on only two of these five: Halo 1 (in the notation of \citealt{Greif11}), which  in the original calculation rapidly fragments into a multiple system, and Halo 2, which undergoes a phase of HD-dominated cooling, which ultimately leads to less fragmentation \citep[c.f.][]{Clark11a}. Halo 1 has a mass of 1810 \msun within its central 2~pc and we denote it in our study as H1. Halo 2 has a mass of 1240 \msun within its central 2~pc, and will be referred to as H2. Halo 1 has only $\sim 6.9\E^5$ SPH particles and Halo 2 $\sim 6.3\E^5$ particles, but due to the on-the-fly refinement used in {\sc arepo} the particle mass in the central regions of the halo is only $10^{-4}$ \msun as mentioned above. We did not include any traditional dark matter particles, but instead treated the dark matter analytically as described in the previous section.

For both halos, we run one simulation in which the effects of DMA are not included (H1-ref, H2-ref) and a second which assumes a DM particle mass of 100~GeV (H1, H2). For Halo 1, we also run two simulations with different values of $m_{x}$: one in which we set $m_{x} = 10$~GeV (H1-lm) and a second in which we set $m_{x} = 1000$~GeV (H1-hm). As the power produced per unit volume by DMA scales as $m_{x} n_{x}^{2} \propto m_{x} \rho_{x}^{2} / m_{x}^{2} \propto \rho_{x}^{2} / m_{x}$, these two runs correspond to cases in which the energy input rate is increased or decreased by an order of magnitude, respectively. As astrophysical constraints strongly disfavour DM masses smaller than $m_{x} = 10 \: {\rm GeV}$, \citep{Schleicher09b,Ackermann11,Galli11}, our H1-lm model should give an indication of the largest effect that we can reasonably expect to obtain from DMA. Details of our simulations are summarised in Table~\ref{sims}.

\section{Collapse to near stellar densities}\label{results1}
\subsection{Densities and temperatures}

\begin{figure*}
\begin{center}
\includegraphics[width=7in]{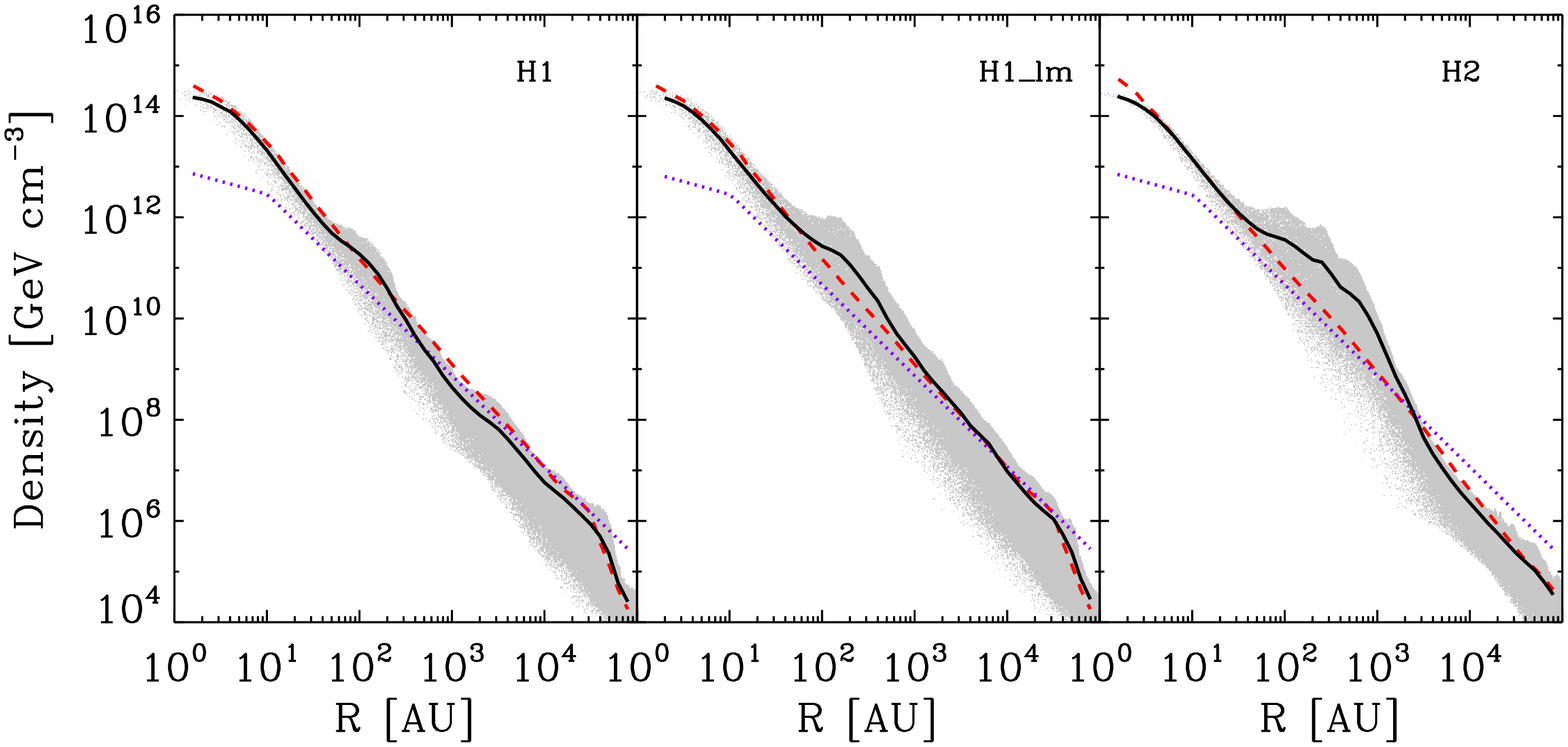}
\caption{Density profiles of the gas and the DM at the point at which the hydrogen nuclei number density of the gas at the centre of the collapsing core first reached
$5 \times 10^{14} \: {\rm cm^{-3}}$. The solid line shows the radially-averaged baryon density and the dotted line shows the DM density. The background grey scale shows the baryon density profile before averaging. The dashed line shows the radially-averaged baryon density in the reference case without DMA. The main effect of DMA is to create an enhancement in the density profile in the region between 100~AU and 1000~AU.}
\label{density}
\end{center}
\end{figure*}

\begin{figure*}
\begin{center}
\includegraphics[width=7in]{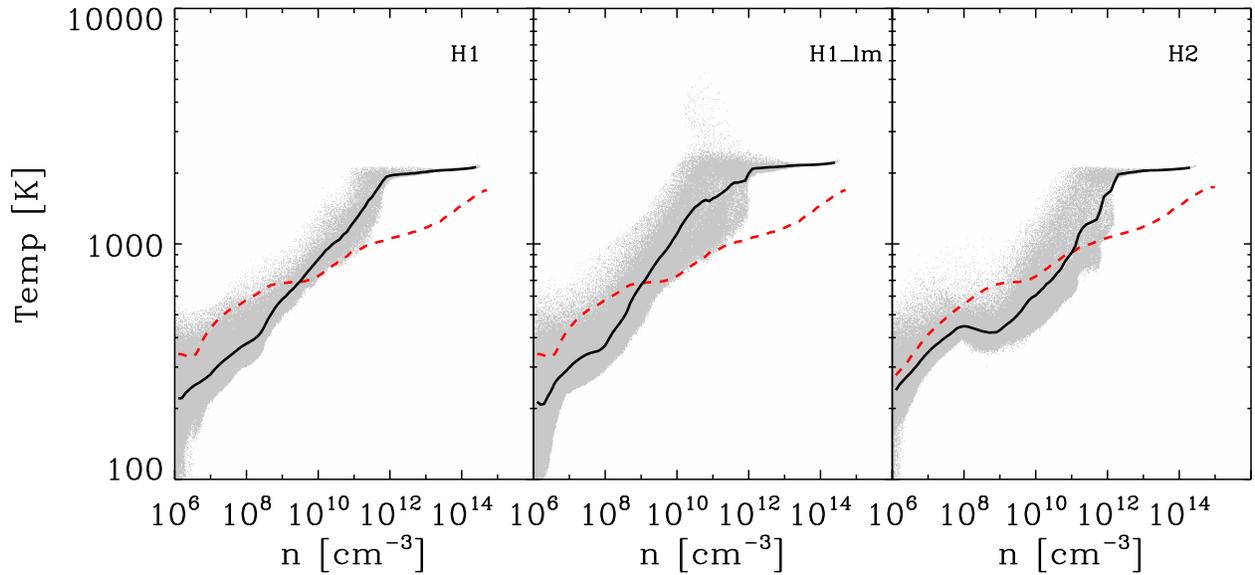}
\caption{Temperature of the gas as a function of density, plotted for the time when the  hydrogen nuclei number density of the gas at the 
centre of the core first reached $5 \times 10^{14} \: {\rm cm^{-3}}$. The solid line shows the radially-averaged gas temperature at each density and the background grey scale shows the gas temperature before averaging. The dashed line shows the radially-averaged gas temperature in the reference case without DMA. With DMA, the gas is cooler in the outer, less dense regions of the halo, but much hotter in the dense interior.}
\label{temp}
\end{center}
\end{figure*}

\begin{table}
	\caption{Time taken to form first sink}
%	collapse from a central density of $10^6$ \cmc to densities sufficient to form a sink particle.}
	\centering
		\begin{tabular}{l c}
   	         \hline
	         \hline
	         Simulation & Time ($10^5$~yr) \\
	         \hline
	         H1-ref & $3.57$ \\
		 H1 & $2.38$ \\
		 H1-lm & $2.70$\\
		 H1-hm & $2.57$\\
		 \hline
		 H2-ref & $3.59$ \\
		H2 & $2.36$ \\
		\hline
		\end{tabular}
	\label{time_c}
\end{table}

In \fig \ref{density}, we show the density profiles of gas and DM for simulations H1, H1-lm, and H2, plus the two reference models. The simulations are compared when the hydrogen nuclei number density at the centre of the halo first reaches $5 \times 10^{14} \: {\rm cm^{-3}}$, which corresponds to a dark matter core radius of $r_c\sim8$ AU and a time of roughly 6 years before the formation of the first protostar. It is immediately apparent that the gas in each halo is able to collapse to high densities, regardless of the strength of the DMA feedback. This is true even in our maximal feedback model, H1-lm, where the DM particle mass was only 10~GeV. For reference, in \citet{Spolyar08} it was predicted that collapse would stop at densities of $10^{13}$ \cmc\ for a DM mass of 100~GeV, and at densities of $10^9$ \cmc\ for a DM mass of 10~GeV. We find no evidence for this in our simulations. On the other hand, our results are in good agreement with the 1D results of \citet{Ripamonti10}, who found that the gas could collapse to densities of $10^{13}$--$10^{14} \: {\rm cm^{-3}}$ in all of their models. We also found collapse up to high densities in run H1-hm, performed with a DM particle mass of 1000~GeV, but as the effects of DMA in this case are much smaller than in our fiducial case, we do not discuss this run further in this section.

The main effect that DMA has on the density profile of the gas is the clear density enhancement at radii of 100--1000~AU. We also find that in the runs with DMA, the time taken for the gas to collapse to protostellar densities is shorter than in the reference runs (see Table~\ref{time_c}). Both of these effects can be understood by an analysis of the temperature structure of the gas. In \fig \ref{temp}, we show how the temperature of the gas varies as a function of the gas number density. In the outer regions of the halo, the gas is cooler than in the reference case. At higher densities, however, there is a sharp rise in the gas temperature, taking it above the value in the reference case. Comparison of Figures~\ref{density} and \ref{temp} shows that the density at which this temperature increase occurs is the same as the density at which we first see the bump in the density profile.

\begin{figure*}
\begin{center}
\begin{tabular}{ c c}
\includegraphics[width=2.5in]{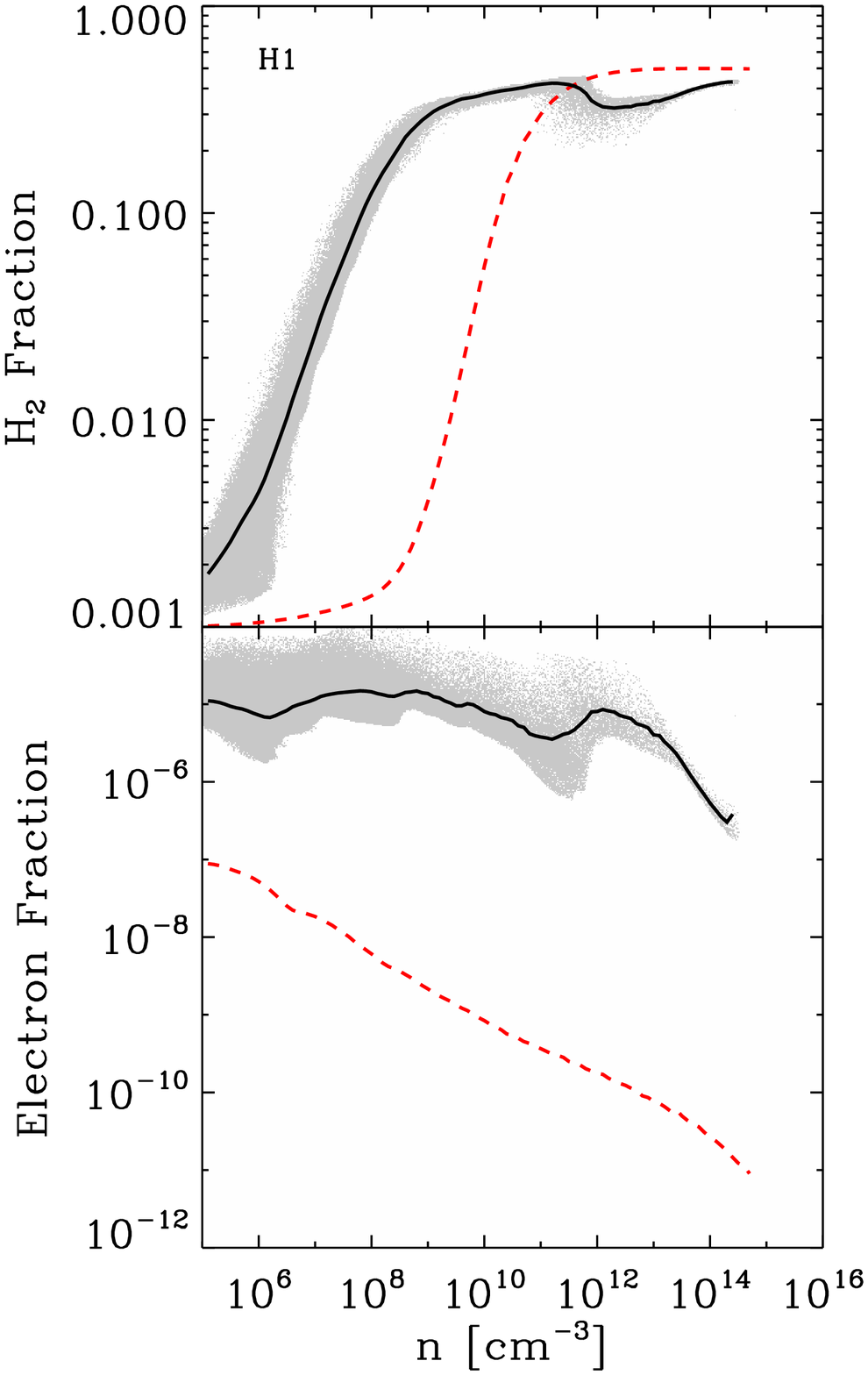}
\includegraphics[width=2.5in]{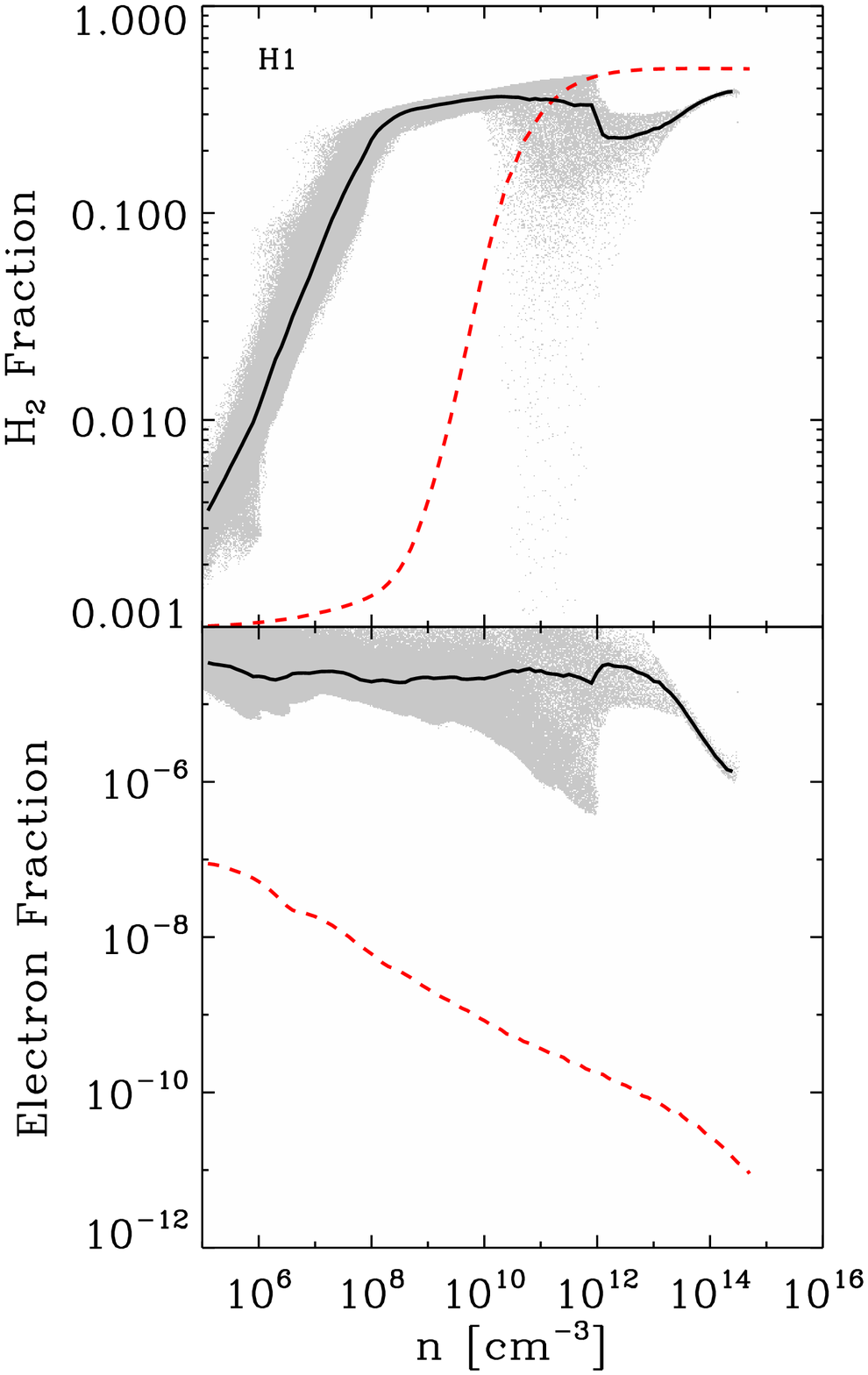}\\
\end{tabular}
\caption{Fractional abundances of H$_2$ and electrons in Halo~1 in the fiducial case (run H1; \textit{left panel}) and the 10 GeV DM particle mass case (run H1-lm; \textit{right panel}). The solid lines show the radially-averaged fractional abundances at each density and the background grey scale shows the abundances before averaging. The dashed lines show the radially-averaged abundances in the reference case without DMA. DMA-induced ionisation increases the electron abundance of the gas at all densities, which in turn leads to an increase in the rate of H$_{2}$ formation at low densities. At high densities,  there is a drop in the H$_2$ abundance due to DMA-induced dissociation of H$_{2}$.}
\label{abund}
\end{center}
\end{figure*}

In Figure~\ref{abund}, we show how the fractional abundances\footnote{Note that these are defined here as the ratio of the number density of the species of interest to the number density of hydrogen nuclei. This means that gas in which all of the hydrogen is in molecular form has an H$_{2}$ fractional abundance $x_{\rm H_{2}} = 0.5$.} of H$_2$ and e$^-$ vary as a function of density within our different models. At low densities, DMA-induced ionisation enhances the number of electrons by two or more orders of magnitude. This increases the rate of H$_{2}$ formation relative to the case without DMA by increasing the effectiveness
of the following reaction chain:
\begin{eqnarray}
{\rm H}+e^- & \rightarrow & {\rm H}^- + \gamma,  \\
{\rm H}^- + {\rm H} & \rightarrow & {\rm H}_2 +e^{-}.
\end{eqnarray}
As H$_2$ is the main coolant of the gas at these densities, the enhanced H$_{2}$ abundance leads to more efficient cooling. We therefore find, counter-intuitively, that this lower density gas is cooler in the case with DMA than in the case without DMA. This is in agreement with the results of the 1D simulations of \citet{Ripamonti10}, and provides an explanation for the shorter collapse times of the halos in which DMA is occurring, as for the majority of its lifetime, the gas has less thermal support.

Once the gas reaches a density of around $10^{12} \: {\rm cm^{-3}}$, however, the H$_2$ fraction decreases sharply in the runs with DMA. By  comparing this Figure with the temperature distribution shown in Figure~\ref{temp}, we see that this sharp decrease occurs once the gas temperature reaches $T \sim 2000$~K. It occurs because at this temperature, collisional dissociation of H$_{2}$ by the reactions
\begin{eqnarray}
{\rm H_{2}} + {\rm H} & \rightarrow & {\rm H} + {\rm H} + {\rm H},  \label{colldiss1} \\ 
{\rm H_{2}} + {\rm H_{2}} & \rightarrow & {\rm H} + {\rm H} + {\rm H_{2}}, \label{colldiss2} \\ 
{\rm H_{2}} + {\rm He} & \rightarrow & {\rm H} + {\rm H} + {\rm He}, \label{colldiss3}
\end{eqnarray}
and by the charge transfer reaction
\begin{equation}
{\rm H_{2}} + {\rm H^{+}} \rightarrow {\rm H_{2}^{+}} + {\rm H},  \label{ct}
\end{equation}
becomes effective. The increased temperature leads to an increased thermal pressure, which slows the collapse of the gas at these densities and leads to a ``pile-up'' of material visible as a pronounced bump in the gas density profile. It is also clear from Figure~\ref{abund} that not all of the H$_{2}$ is destroyed at this density. The H$_{2}$ that survives in the higher density gas enables it to dissipate much of the energy introduced into the gas by the DMA, and hence allows it to maintain its temperature at close to 2000~K. This allows gravitational collapse to continue past the point at which DMA heating first outstrips H$_{2}$ cooling. We examine the physics of this in more detail in the next
section.

\begin{figure}
\begin{center}
\includegraphics[width=3in]{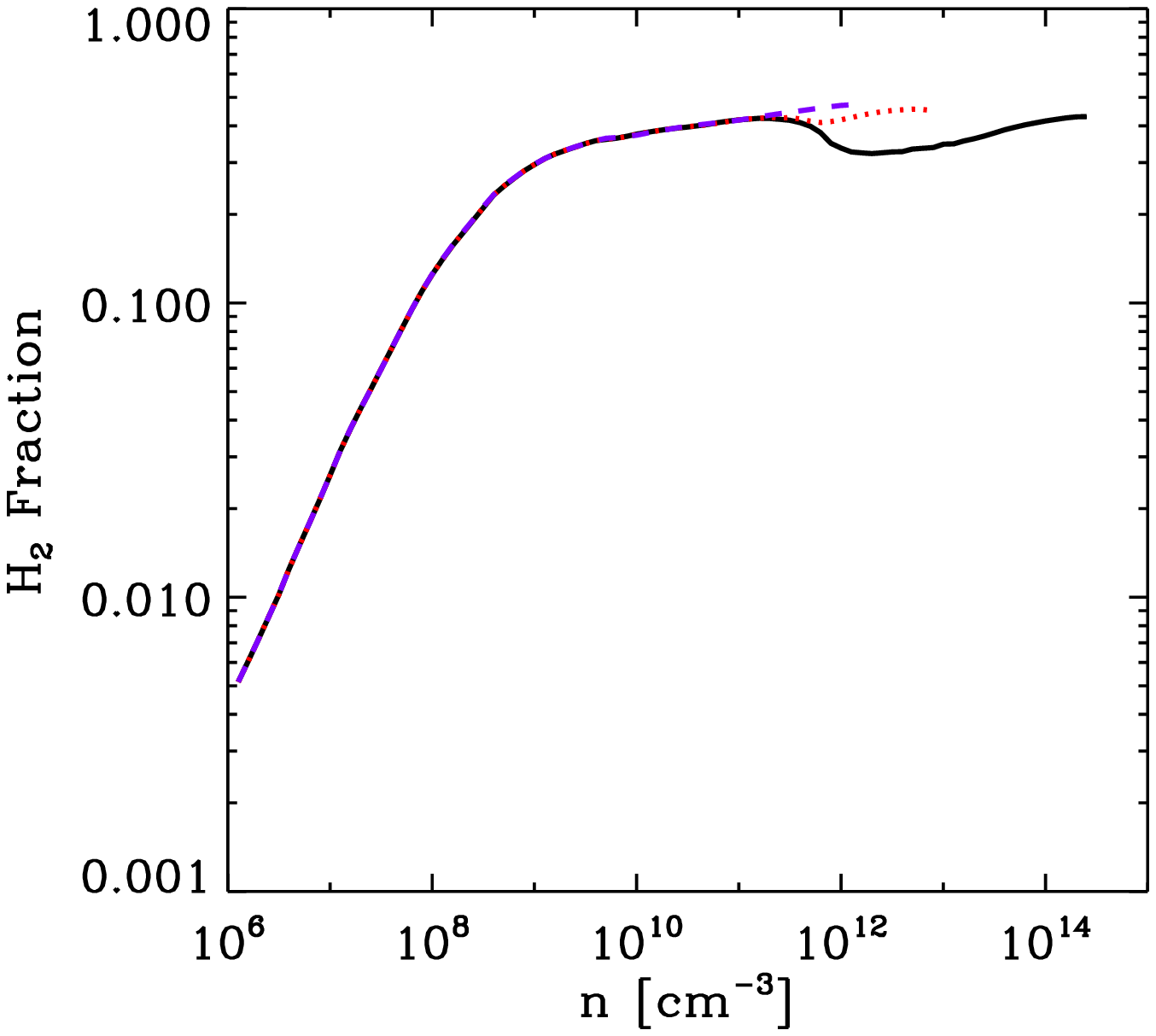}
\caption{Radially-averaged fractional abundance of H$_{2}$ as a function of density in run H1 at three different output times, corresponding to peak densities of approximately $10^{12} \: {\rm cm^{-3}}$ (dashed line), $10^{13} \: {\rm cm^{-3}}$ (dotted line) and $5 \times 10^{14} \: {\rm cm^{-3}}$ (solid line). We see that as time goes on, the gas loses an increasingly large fraction of its H$_{2}$ content once it reaches $n \sim 10^{12} \: {\rm cm^{-3}}$. The increase in the H$_{2}$ fraction at higher densities is therefore a consequence of the fact that gas which collapsed earlier lost less H$_{2}$ than gas which collapsed later, and does not indicate reformation of H$_{2}$ within the densest gas.}
\label{abund-time}
\end{center}
\end{figure}

Finally, we note that one must take care in interpreting the behaviour of the H$_{2}$ abundances shown in Figure~\ref{abund}. It is all too easy to think of the evolution of $x_{\rm H_{2}}$ with density as a time history of the H$_{2}$ abundance. In this picture, the H$_{2}$ abundance first increases as the gas collapses, then sharply decreases once the number density reaches $10^{12} \: {\rm cm^{-3}}$, before increasing once more at higher gas densities. However, this is incorrect, as Figure~\ref{abund-time} demonstrates. The rise in the H$_{2}$ abundance in the densest gas that we see if we look at a single snapshot is not due to the reformation of H$_{2}$ in this gas. Instead, it occurs because the amount of H$_{2}$ that the gas loses once it reaches a density of around $10^{12} \: {\rm cm^{-3}}$ increases as the collapse proceeds. The first gas to collapse -- the material that is now at densities above $10^{14} \: {\rm cm^{-3}}$ -- loses only a relatively small fraction of its H$_2$, while the material falling in later loses much more of its molecular content.  In part, this behaviour is due to an increase in the DM number density associated with the $n = 10^{12} \: {\rm cm^{-3}}$ gas, caused by the increasing concentration of the halo. However, an additional contribution comes from the fact that gas falling in at later times will generally have a higher infall velocity, and hence will shock more strongly once it reaches this density. 

\subsection{Heating and cooling rates}
\subsubsection{Fiducial case}

\begin{figure*}
\begin{center}
\includegraphics[width=4.5in]{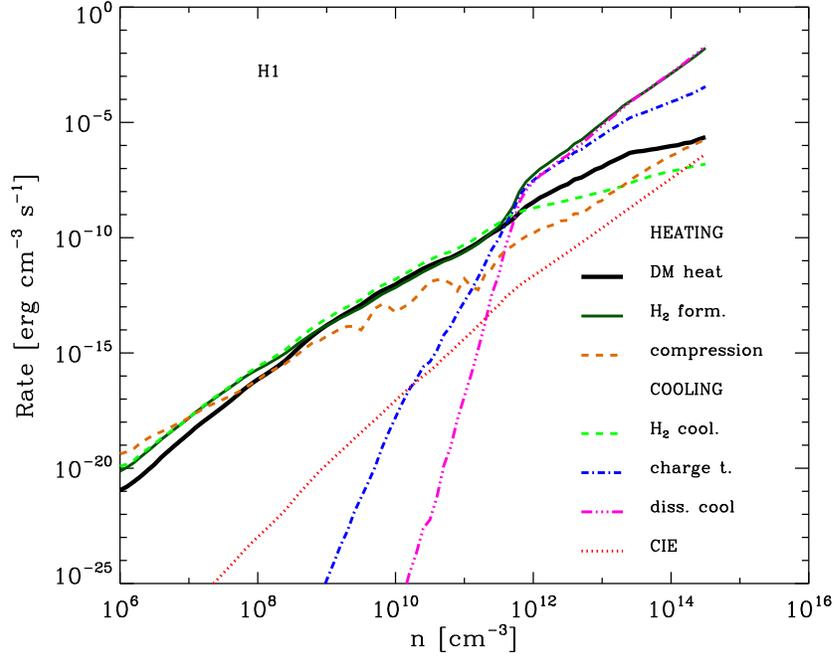}
\caption{Rates of the major heating and cooling processes active in run H1 at the time when the central density first reaches $n = 5 \times 10^{14} \: {\rm cm^{-3}}$. DMA heating is an important contributor to the net heating between densities of $10^8$--$10^{12} \: {\rm cm^{-3}}$. However, over most of this range, it is largely balanced by H$_{2}$ line cooling, and provides insufficient heating to halt the collapse. Above $n \sim 10^{12} \: {\rm cm^{-3}}$, H$_{2}$ line cooling becomes ineffective, and most of the energy introduced into the gas by DMA heating is dissipated by H$_{2}$ collisional dissociation and the destruction of H$_{2}$ by charge transfer.}
\label{rates_stand1}
\end{center}
\end{figure*}

\fig \ref{rates_stand1} shows the main heating and cooling processes acting in run H1 shortly before the formation of the first protostar, at the time when the central number density first reached $5 \times 10^{14} \: {\rm cm^{-3}}$. At low densities, compressional heating and H$_{2}$ formation heating are the two most important heat sources, although even at densities as low as $10^{6} \: {\rm cm^{-3}}$, DMA heating is beginning to contribute to the total heating rate at the level of a few percent or more. The cooling of the gas at these densities is dominated by H$_{2}$ line cooling. Once the gas density reaches $n \sim 10^{8} \: {\rm cm^{-3}}$, compressional heating starts to become unimportant, and DMA heating catches up with H$_{2}$ formation heating, with both subsequently playing important roles in the thermal balance of the gas. The heating produced by these two processes raises the gas temperature, but this enables H$_{2}$ line cooling to become more effective, and this is able to offset much of the additional heat input from the DMA at densities $n < 10^{12} \: {\rm cm^{-3}}$. Above this density, H$_{2}$ line cooling becomes relatively ineffective, both because the H$_{2}$ is starting to dissociate and also because the H$_{2}$ emission lines become optically thick at high densities. The heating rate due to DMA therefore outstrips the H$_{2}$ line cooling rate, in agreement with the prediction of \citet{Spolyar08}. However, we see from Figure~\ref{rates_stand1} that this is not the whole story. As the temperature rises, H$_{2}$ collisional dissociation (reactions~\ref{colldiss1}--\ref{colldiss3}) and the destruction of H$_{2}$ by charge transfer with H$^{+}$ (reaction~\ref{ct}) become increasingly effective. These reactions are endothermic, and hence remove thermal energy from the gas. If H$_{2}$ subsequently reforms, then the associated H$_{2}$ formation heating will return some of this energy to the gas, but if more H$_{2}$ is destroyed than can reform, then the overall effect is to dissipate energy. This effect should not be regarded as ``cooling'' in the same sense as H$_{2}$ line cooling or CIE cooling, as it will not bring about a decrease in the gas temperature. Instead, it acts more like a thermostat, preventing the temperature from increasing significantly until all of the H$_{2}$ has been consumed. 

The importance of this effect can be appreciated if we compare the time required to destroy all of the H$_{2}$ at the centre of the halo with the free-fall collapse time of the gas. By considering the energy budget of the gas in this way we can obtain a rough estimate of whether collapse is likely to be halted at densities close to our resolution limit. From Figure~\ref{rates_stand1}, we see that at a density $n = 5 \times 10^{14} \: {\rm cm^{-3}}$,  the DMA heating rate per unit volume at the centre of the halo is roughly $5 \times 10^{-6} \: {\rm erg} \: {\rm s^{-1}} \: {\rm cm^{-3}}$. If we take the H$_{2}$ fractional abundance in this gas to be $x_{\rm H_{2}} = 0.4$, which is a conservative estimate, then the total amount of energy per unit volume that can be dissipated by dissociating H$_{2}$ is
\begin{eqnarray}
E_{\rm H_{2}} & = & 4.48 {\rm eV} \times x_{\rm H_{2}} n, \\
 & \simeq & 1400 \left(\frac{n}{5 \times 10^{14} \: {\rm cm^{-3}}}\right) \: {\rm erg} \: {\rm cm^{-3}},   
\end{eqnarray}
where $4.48 \: {\rm eV}$ is the binding energy of a single H$_{2}$ molecule. The time required to destroy all of the H$_{2}$ is therefore $t_{\rm dis} = 1400 / 5 \times 10^{-6} \sim 3 \times 10^{8} \: {\rm s}$,  or around 10~years. In comparison, the free-fall time of the gas at this density is around 2~years. The energy input from the DM therefore cannot destroy the H$_{2}$ rapidly enough to prevent the gas from collapsing. At even higher densities, the DMA heating rate will be larger. However, the heating rate per unit volume within the core of the DM density profile scales with the central gas density as $n^{1.62}$, and hence $t_{\rm dis} \propto n^{-0.62}$. In comparison, the free-fall timescale scales as $t_{\rm ff} \propto n^{-0.5}$. Therefore, a large increase in density is required in order to significantly alter the ratio of the H$_{2}$ dissociation timescale to the free-fall timescale.

\begin{figure*}
\begin{center}
\includegraphics[width=4.5in]{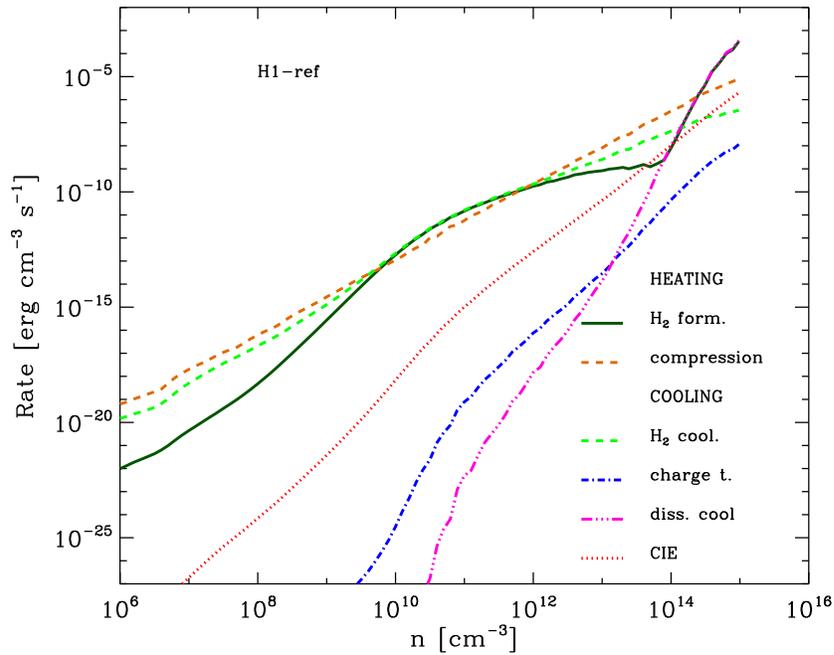}
\caption{Rates for the major heating and cooling processes acting in simulation H1-ref at a point just before the formation of the first protostar. We show the rates at a later time than in Figure~\ref{rates_stand1} so that the behaviour in the high density regime dominated by H$_{2}$ dissociation cooling is clear. If we compare the results here with those in Figure~\ref{rates_stand1}, we see that H$_{2}$ dissociation cooling only becomes important at $n \sim 10^{14} \: {\rm cm^{-3}}$, in contrast to $n \sim 10^{11} \: {\rm cm^{-3}}$ in run H1. This is because the gas temperature at these densities is lower in run H1-ref than in run H1, owing to the absence of DMA heating.}
\label{rates_ref1}
\end{center}
\end{figure*}

In reality collapse of the dense gas  is likely to occur on a timescale that could be a factor of a few longer than the free fall timescale. However, even in this case the collapse timescale is shorter than the H$_2$ dissociation timescale, and will remain so until the density increases significantly. It therefore seems unlikely that collapse will halt at a density just above our sink creation threshold, although we cannot say at exactly what point the collapse will halt without performing much higher resolution simulations.

Figure~\ref{rates_ref1} shows the main heating and cooling processes acting in the reference case, run H1-ref. In the absence of DMA heating, the main heating term is compressional heating over most of the range of densities examined, with H$_{2}$ formation heating becoming important at densities between $10^{10} \: {\rm cm^{-3}}$ and $10^{12} \: {\rm cm^{-3}}$, and at $n > 10^{14} \: {\rm cm^{-3}}$. The main source of cooling at $n < 10^{14} \: {\rm cm^{-3}}$ is H$_{2}$ line cooling, while at higher densities, H$_{2}$ collisional dissociation plays an important role in regulating the temperature of the gas. This is in reasonable agreement with the results of other models; for instance, \citet{Yoshida06} find that H$_2$ dissociation becomes significant at densities of $\sim10^{15} \: {\rm cm^{-3}}$, at which point the gas temperature is roughly 2000~K. The main effect of the DMA heating seems to be simply to bring the gas to this state at an earlier point in its evolution.

\subsubsection{Maximal Case}
\begin{figure*}
\begin{center}
\includegraphics[width=4.5in]{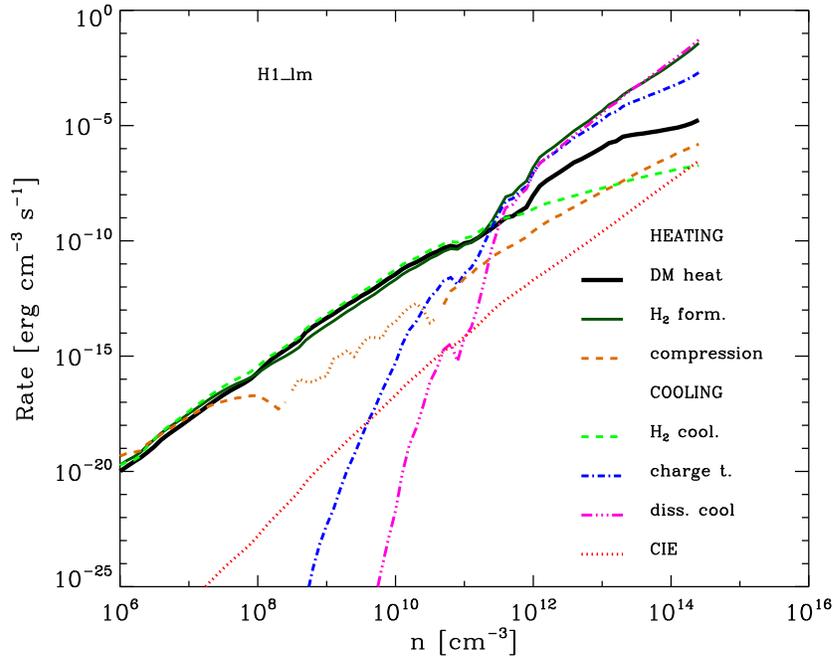}
\caption{As \fig \ref{rates_stand1}, but for run H1-lm. In this case, the heating from DMA is important at all densities.}
\label{rates_lm1}
\end{center}
\end{figure*}

The previous analysis is for the fiducial case where the DM particle mass was 100~GeV. However, even in our maximal case, where the DM particle mass was 10~GeV, we find broadly similar behaviour. In Figure~\ref{rates_lm1}, we show the rates of the main heating and cooling processes in run H1-lm shortly before the formation of the first protostar. In this case, we see that DMA heating becomes important at an earlier time and that H$_{2}$ dissociation becomes the dominant energy dissipation mechanism once the gas density reaches a lower density, $n \sim 10^{10} \: {\rm cm^{-3}}$, than in the fiducial model. This is consistent with the results plotted in Figure~\ref{temp}, which show that the gas reaches $T \sim 2000$~K slightly earlier in its evolution. However, the subsequent behaviour of the gas is very similar in both models. H$_{2}$ dissociation is such an effective thermostat that an increase in the heating rate by a factor of ten produces only a small increase in the gas temperature. Repeating our previous analysis of the H$_{2}$ dissociation timescale, we find that in this case $t_{\rm dis} \sim t_{\rm ff}$, meaning that the gas will probably lose most of its H$_{2}$ before it collapses to protostellar densities. It is therefore possible that in this extreme case, a ``dark star'' will form, but if so, then its size will be smaller than our minimum spatial resolution of a few AU. For comparison, \citet{Spolyar08} predict that even in the 100~GeV case, a dark star of size $\sim 20$~AU should form, and expect that reducing the DM particle mass will lead to an even larger dark star.  We find no evidence for such large DMA supported structures in our simulations.

\section{Secondary Fragmentation}\label{results2}
So far we have only considered the initial collapse of the halo to stellar densities. However, \citet{Stacy10} and 
\citet{Clark11b} have shown that  the protostellar accretion disc that builds up around the first protostar rapidly 
becomes unstable and fragments. The natural implication is that the first stars to form were generally part of multiple systems, which has profound consequences for our understanding of primordial star formation. It raises the possibility of the first stars being ejected while they were still low mass \citep{Greif11} or growing in mass through mergers \citep{Smith12b,Greif12}. Evolving close binary systems are also a potential mechanism for creating early gamma ray bursts \citep{Bromm06}. It is therefore important to discover whether primordial accretion discs are still unstable in the presence of DMA.

In order to enable us to follow the evolution of our simulated halos beyond the point at which the first protostar forms, we use a sink particle treatment, as outlined in Section~\ref{basic}. During the evolution of the halo with sink particles we fix the location of the DM peak to exactly coincide with the first sink formed. This mimics the effect of the DM being `locked in' to the first star to form.
%Sinks are created only once the gas density exceeds $10^{16} \: {\rm cm^{-3}}$ and only in gas which is gravitationally bound and collapsing, without the possibility of re-expansion. Our sink particles have an outer accretion radius of 6 AU, within which they accrete all gravitationally bound material, and an inner accretion radius of 4 AU, within which every SPH particle is accreted. We also include  accretion luminosity feedback, as described in \citet{Smith11b}, in order to ensure that the effects of feedback are maximal. During the evolution of the halo with sink particles we fix the location of the DM peak to exactly coincide with the first sink formed. This mimics the effect of the DM being `locked in' to the first star to form.

\begin{figure*}
\begin{center}
\includegraphics[width=7in]{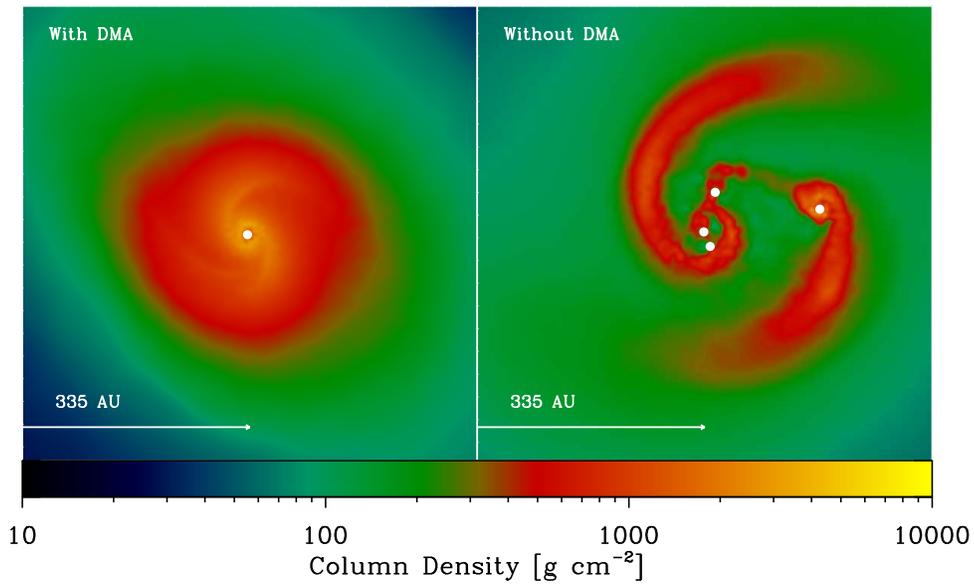}
\caption{Column density of a slice through the midplane of the disc in the central regions of runs H1 and H1-ref at a time 500 years after the first sink formed in each case. The slice is 1000 AU thick. In the reference case without DMA, four fragments are formed. In the DMA case, there is no secondary fragmentation even at later stages of the evolution.}
\label{frag_2panel}
\end{center}
\end{figure*}

\fig \ref{frag_2panel} shows a comparison of the column density of the disc in run H1 and in the corresponding reference case without DMA, run H1-ref, at a time 500~years after the formation of the first protostar. In the reference case, the disc has already fragmented, forming four additional protostars. In the case with DMA heating, however, the disc has not fragmented. We continued run H1 until the primary mass reached $15 \: {\rm M_{\odot}}$,  which occurred at a time $t = 17,527$~years after the formation of the first protostar, but saw no disc fragmentation during the whole of this period.  As our models do not include the effects of ionisation feedback from massive stars, we stopped our simulation at this point. However, we note that studies that have investigated the effect of this ionisation feedback find that it rapidly shuts off the supply of material to the disc, and hence will act to suppress fragmentation at later times \citep{Stacy12,Hosokawa11b}.

\begin{figure*}
\begin{center}
\includegraphics[width=4.5in]{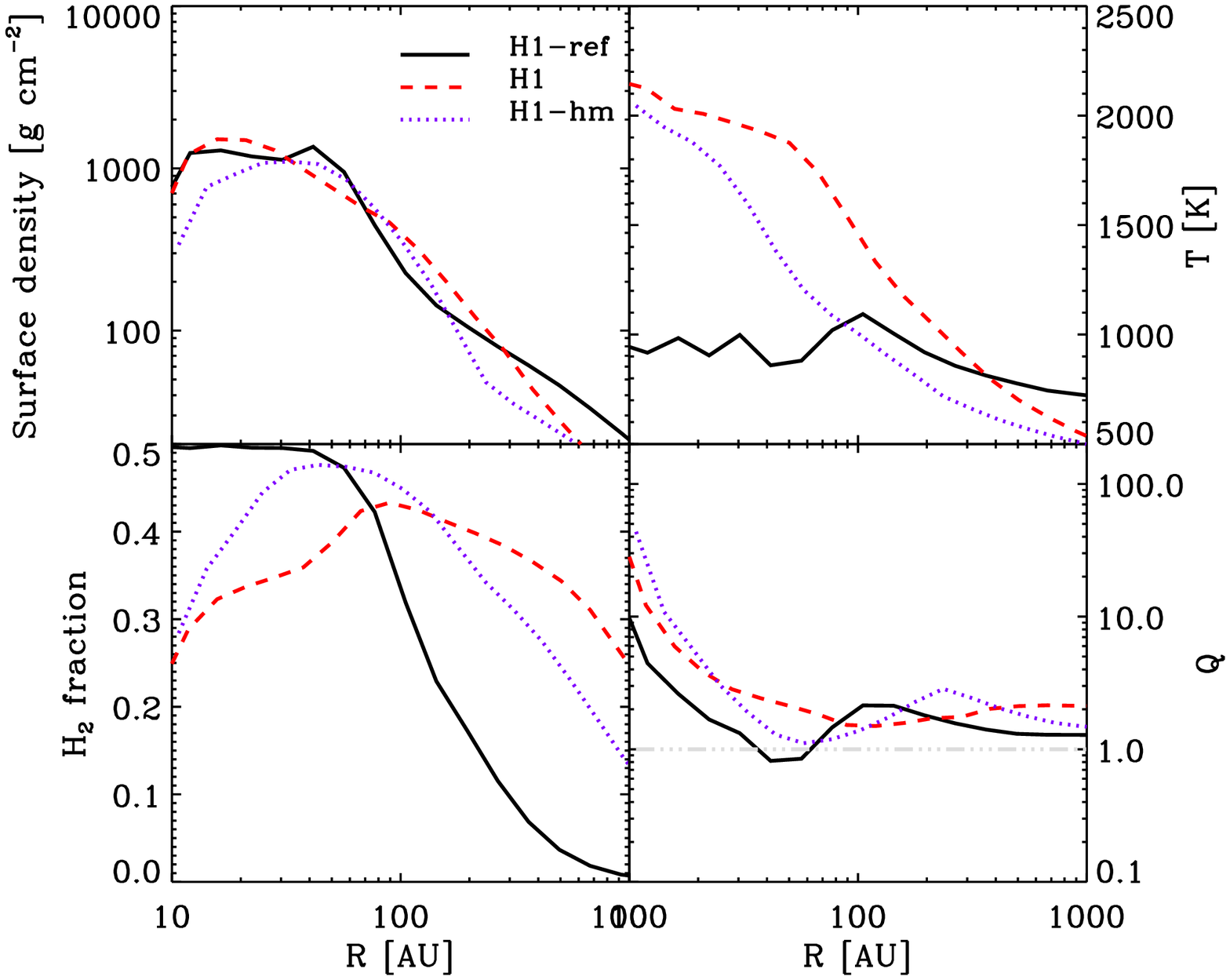}
\caption{Properties of the disc 200 years after the first sink forms. The solid line shows the reference case H1-ref, the dashed line our fiducial case H1, and the dotted line shows our minimal case H1-hm. In the reference case, the disc has become Toomre unstable, but in the two cases with DMA, the disc remains stable. Note that the low DM mass case, H1-lm does fragment at radii of order 1,000 AU later in the simulation.}
\label{Q}
\end{center}
\end{figure*}

An indication of disc stability is given by the Toomre Q parameter \citep{Toomre64}, which has the form
\begin{equation}
Q= \frac{c_s \kappa}{\pi G \Sigma}
\end{equation}
,where $c_s$ is the sound speed, $\kappa$ is the epicyclic frequency, and $\Sigma$ is the surface density of the disc. In the classical Toomre analysis, discs are stable if they have $Q > 1$.  Such an analysis is technically only valid when the disc mass is much smaller than that of the central protostar, and the disc is infinitely thin, neither of which are the case here. However, $Q$ still proves to be a reliable guide as to whether or not the disc is stable, and helps to demonstrate why the inclusion of DMA heating has a large effect on the disc stability. \fig \ref{Q} shows the radially averaged properties of the disc surrounding the central protostar  for Halo 1. We plot three cases, the reference case without DMA (H1-ref), the fiducial case with DMA (H1), and the ``minimal'' case with a high DM particle mass (H1-hm). The comparison between the disc properties is made at a time of 217~yr after the first sink particle forms in each simulation. For the reference case, this corresponds to a time just before disc fragmentation. (The model of low DM mass, H1-lm, shall be discussed in more detail later.)

All of our discs rotate at a similar rate, for instance H1 has a mean radial velocity of 4.9 \kms\ in the gas denser than $1\E^{10}$ \cmc\ and H1-ref has a mean radial velocity of 4.2 \kms in the gas at this density. Consequently, the most relevant terms are the surface density of the disc and the sound speed. \fig \ref{Q} shows that the surface density of the disc is not substantially changed by the DMA. However, the disc temperature is considerably increased in the DM case compared to the reference case. This raises the sound speed and causes the Q value to rise above 1 in the DM cases, whereas in the reference case it falls below one and the disc fragments. \fig \ref{Q} also shows the H$_2$ fraction within the disc. In the reference case, the disc is fully molecular, but in the DM cases the disc is partially dissociated. As the disc evolves the H$_2$ fraction further decreases until all the molecular gas is destroyed. After this point the energy can no longer be absorbed by the destruction of H$_2$ and the temperature of the disc will rapidly rise. Thus it becomes increasingly difficult for the disc to fragment as it evolves, which explains why we see no fragmentation in run H1 despite running the simulation to a central primary mass of $15 \: {\rm M_{\odot}}$.

\begin{figure*}
\begin{center}
\includegraphics[width=7in]{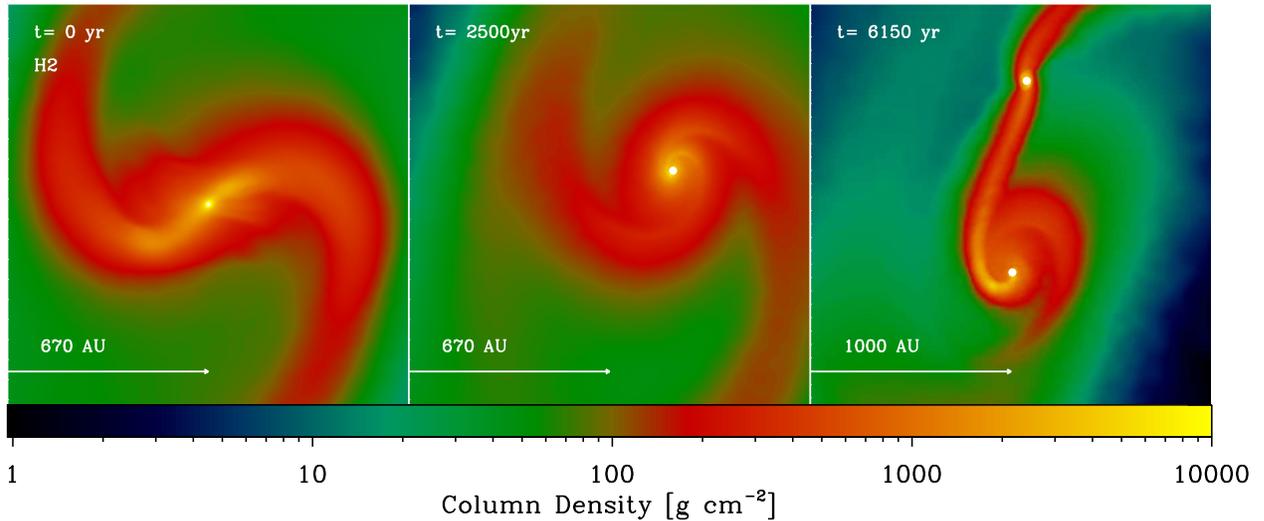}
\caption{Column density projections of a slice through the midplane of the disc in the central regions of run H2. The slice is 1000 AU thick. In this case a secondary protostar is formed, but at a very large separation from the central protostar. In the second panel, the image is centred on the original position of the first protostar. It is currently unclear how the movement of the protostar after its formation will affect the DM profile.}
\label{H2frag}
\end{center}
\end{figure*}

\fig \ref{H2frag} shows a column density projection for a slice through the centre of run H2 at three different output times. In this case, a second sink forms at large radii where the gas is cooler. The fragmentation in run H2 is substantially different from that seen in the reference model. Without DMA, a secondary sink forms within 155 years at a distance of 14.8 AU from the primary. This is followed by additional fragmentation such that when the simulation has run for a thousand years, there are 7 sinks. In contrast to this, the DM simulation does not form a secondary object until $t = 6090 \: {\rm years}$, and the fragmentation occurs at a distance of 1012 AU from the primary. The two stars then form a wide binary system. 

\begin{table}
	\caption{Protostellar separation and primary mass at the point when a secondary protostar forms}
	\centering
		\begin{tabular}{l c c}
   	         \hline
	         \hline
	         Simulation & Separation [AU] & Primary Mass [${\rm M_{\odot}}$] \\
	         \hline
	         H1-ref & 31.7 & 1.63\\
		 H1 & - & -\\
		 H1-lm & 1629 & 11.8\\
		 H1-hm & - & -\\
		 \hline
		 H2-ref & 14.8 & 0.68 \\
		H2 & 1012 & 10.3 \\
		\hline
		\end{tabular}
	\label{secondary}
\end{table}

\tab \ref{secondary} summarises the secondary fragmentation in all of our simulations. It is interesting that secondary fragmentation occurs in run H2 but not in run H1 for our fiducial DM mass case. \fig \ref{density} shows that in run H2, a larger bump forms in the density profile than in run H1. Similarly, there is also secondary fragmentation in H1-lm at distances of around a thousand AU where there was a significant enhancement in the density profile. At such large radii the gas is far enough away from the peak DM density that DMA cannot dissociate H$_2$ and the gas remains fully molecular, allowing it to cool effectively. Runs H1-lm and H2 both had higher DMA levels than H1 and H1-hm and consequently a greater density enhancement outside the DM core. (Our normalisation of the DM in Halo 2 meant that the ratio of baryonic matter to DM was slightly lower in this case.) Consequently it can be concluded that DMA suppresses fragmentation in the disc close to the primary but it can actually encourage fragmentation in spiral arms at distances of a few thousand AU.

Another striking difference between the cases with and without DMA is the mass of the primary at the time that the disc first fragments. In the reference case, the primary is a low mass object which is evolving adiabatically. Its internal structure is that of an extended ball of gas \citep{Omukai03} and it will be prone to interactions with close neighbours. With DMA, however, the primary is around ten solar masses at the point of fragmentation, at which point it undergoes an expansion as it redistributes its internal entropy and will shortly thereafter begin contracting to the main sequence. The later evolutionary stage of the primary and greater distance from the secondary in the DMA case mean it is unlikely that the stars will influence each other significantly during their formation.

The left and middle panels of \fig \ref{H2frag} are centred on the original position at which the sink particle is originally formed. It can be seen that there is some drift in the position of the sink after the first few thousand years. As our DM halo is analytic and cannot dynamically respond to the baryons we assume that the DM is locked into the first protostar and follows this centre as its position migrates. Similarly our idealised DM halo is oblivious to the formation of spiral arms in the baryonic component of the halo. It is conceivable that both the drift of the sink particle and the transfer of energy from the spiral arms to the DM distribution could heat the central DM distribution and somewhat decrease the DM central density. Studies with a highly resolved live DM halo and high resolution baryonic component would be needed to fully address this issue.

\section{Discussion}\label{discussion}
\subsection{Do dark stars form?}
In their original study, \citet{Spolyar08} made the simple assumption that the gravitational collapse of the gas would come to a halt as soon as the heating rate produced by DMA exceeded the H$_{2}$ cooling rate. As a result, they predicted the formation of large ($\sim 20$~AU or more) protostellar objects supported by DMA heating, the so-called ``dark stars''. However, their assumption is incorrect. \citet{Ripamonti10} first showed, using 1D spherically-symmetric models, that the collapse does not halt once the DM heating rate exceeds the H$_{2}$ cooling rate, and in our present study we confirm their results. The key factor that allows collapse to continue is the collisional dissociation of H$_{2}$. This acts as a thermostat, preventing the gas temperature from increasing significantly above 2000~K until all of the H$_{2}$ is destroyed. A simple estimate of the timescale on which the H$_{2}$ is destroyed shows that even in the most extreme case, it is comparable to the free-fall time of the gas, meaning that the gas can reach much higher densities (and hence smaller length scales) than anticipated in the \citet{Spolyar08} study. As a result, we find no evidence for the formation of dark stars within our simulations, although we note that we cannot rule out the formation of such objects if they form above densities of $10^{16}$ \cmc, have initial masses of less than 0.01 \msun and radii of less than an AU, as this is below our resolution limit.

%Our study also demonstrates that even if the first protostar to form is a dark star, it is unlikely to accrete all of the gas falling in to the centre of the halo. Although the accretion disc that forms around the protostar is far more stable than in the case without DMA, it does eventually fragment in several of our simulations, and it is conceivable that it would eventually fragment in every case were we to run the simulations for long enough. We therefore consider the formation of super-massive dark stars of the kind proposed by \citet{Freese10} to be highly unlikely.

\subsection{Stellar Multiplicity}
Perhaps the most striking finding of our work is the large reduction of the level of fragmentation within the protostellar accretion disc. Of our four simulations with DM, secondary fragmentation occurred in only two cases and then only at large radii. Much of the recent focus of studies of Population III star formation has  been on the fragmentation of protostellar discs \citep[see e.g.][]{Clark11b, Greif11,Smith11b}. \citet{Greif12} have shown that fragmentation occurs on scales of 10 AU or smaller, and that mergers between protostars should be common. The possibility of mergers and interactions between protostars was also highlighted by \citet{Smith12b}, who show that the high
accretion rates experienced by the young protostars lead to large, extended, ``fluffy'' objects that have a higher probability of interacting than more conventional protostars.

Such scenarios are far less likely when the effects of DMA are taken into account. In this picture either a single massive star forms at the centre of the halo, or a secondary forms in a wide binary with a separation of around 1,000 AU. Fragmentation on such scales was also seen by \citet{Stacy10} in lower resolution studies which focussed on scales larger than the inner disc around the protostar. In our model, when secondary fragmentation begins the mass of the primary protostar is already greater than 10 solar masses. \citet{McKee08} find that for an accretion rate of $10^{-3}$ \myr\ an ionised HII region will form around the protostar after it obtains a mass of around 20 \msun, depending upon its rotation. Such an ionised region is likely to form even earlier in our model due to the additional ionisation of hydrogen from DMA. Thus it is possible that at the point when secondary stars form the original protostar will already be surrounded by an HII region and any further growth of the primary will be determined by the balance of radiative transfer effects. \citet{Hosokawa12} find that UV radiation from primordial protostars would photo-evaporate their accretion discs once they achieved a mass of around 43 \msun. We note however that similar claims for present day high-mass star formation \citep{Yorke02} have not borne out in detailed 3D calculations \citep{Krumholz09,Peters10,Kuiper12}.

\subsection{Accuracy of DM model}
The biggest assumption in our model is the adoption of an analytic DM halo rather than a live halo. No calculation has yet been able to fully follow the contraction and fragmentation of baryonic matter down to AU scales with a comparable resolution in DM. Simulations by \citet{Abel02} followed the collapse of a minihalo with a mass resolution of 1.1 \msun for the DM component and confirmed that the DM had a peaked profile to radii as small as $\sim 1,500$ AU. Previous work by \citet{Ripamonti10} and \citet{Spolyar08} adopt the method proposed by \citet{Blumenthal86} to model the contraction of a DM halo due to baryonic collapse, and upon these results we base our DM profile. In the Appendix of \citet{Ripamonti10}, the applicability of this method is discussed and found to be in better agreement with the current numerical findings than alternative models \citet[e.g.][]{Steigman78}. We are therefore confident that our model for increasing the DM density with increasing baryon density is reasonable. If anything such a profile may give a slight overestimate of DM density at the centre of the halo. This would further reduce the effective of DMA heating, and make it even less likely that any ``dark stars'' are formed. 

A bigger potential uncertainty in our adopted DM profile, however, is the assumption of spherical symmetry. Before the formation of the first sink particle this is a reasonable simplification, as the baryon profile itself is smooth and centrally concentrated.  However, after the first sink particle is formed, this rapidly changes within the central regions of the halo. A disc quickly develops, and strong spiral arm features develop within the disc. It is possible that interactions between these high asymmetric features and inhomogeneities in the DM density distribution could lead to a transfer of energy to the DM component and a flattening of the central DM distribution, in a similar manner to that thought to occur from a rotating bar in disc galaxies \citep{Weinberg02}. To fully resolve these issues, calculations with a live three-dimensional halo component would be needed.

Another important issue is whether the central protostar remains within the centre of the DM distribution, as this affects the amount of DM that can be captured by the protostar by particle scattering, which strongly influences the lifetime of any dark star phase \citep{Iocco08b,Taoso07,Yoon08}. In our calculations, we find that the central protostar moves by a significant amount after its formation, as a result of the fact that accretion onto the protostar does not occur in a perfectly symmetrical fashion, and also because it is being acted on by torques from the asymmetrical gas distribution in the protostellar accretion disc. We have assumed that the DM responds in a similar fashion, and that the peak in the DM distribution follows the central protostar. However, if an offset were to develop between the DM cusp and the baryon peak, dynamical heating of the DM would occur and the resulting reduction in the DM density would reduce the amount of annihilation. 

This issue has been addressed recently by \citet{Stacy12}, who performed a calculation in which they re-simulated the inner regions of a halo with a live DM halo of similar resolution to that of the gas after the formation of an initial sink particle. The effects of DMA were not, however, included. They formed additional sink particles at around 1,000 AU and found that the motion of the star-disc system became displaced from the DM density peak and that after 5,000 yr there was insufficient DM to influence the formation and evolution of the protostars. However, \citet{Stacy12} do not resolve protostar formation in the inner disc around the central object and do not include DMA and so cannot comment on fragmentation in this regime. In our runs without DMA, secondary fragmentation occurred on timescales of only a few hundred years. It is therefore likely that there will be sufficient DM during the period in which the inner disc is prone to fragmentation to maintain the findings of this paper.

\section{Conclusions}\label{conclusion}
We have for the first time included the effects of dark matter annihilation into 3D simulations of the collapse and fragmentation of cosmological minihalos. We used the SPH code {\sc gadget 2}, which has been modified to include a time-dependent chemical network which includes the effects of dark matter annihilations. The dark matter distribution was modelled analytically based on the predictions of adiabatic contraction \citep{Blumenthal86} and the results of \citet{Spolyar08} and \citet{Ripamonti10}. We adopted a fiducial dark matter particle mass of 100~GeV but also ran simulations with particles masses an order of magnitude larger and smaller to account for uncertainties in our knowledge of the true dark matter particle mass and annihilation cross-section. 

Our main findings are as follows:
\begin{enumerate}
\item Dark matter annihilation does not halt the gravitational collapse of the gas on the scales considered here, contrary to the suggestion made by \citet{Spolyar08}. The reason for this is that collisional dissociation of H$_{2}$ provides an effective way to dissipate large amounts of energy even in the regime where H$_{2}$ line cooling is ineffective, and
hence prevents the gas temperature from increasing much above 2000~K. The timescale for H$_2$ destruction is typically longer than the free-fall time of the gas, even at high
gas densities, and so the central protostar should be able to reach very high densities. This is true even when a dark matter particle mass of only 10~GeV is adopted, which gives 
a maximal effect for dark matter annihilation. It is therefore plausible that a conventional star will form, rather than a ``dark star'', although confirmation of this point awaits a simulation able to follow the collapse of the gas all the way up to protostellar densities.

\item Dark matter annihilation prevents subsequent fragmentation in the disc around the central protostar within a radius 1,000~AU, regardless of whether this central object is a ``dark star'' or a normal protostar. This is primarily due to the dark matter annihilation heating raising the temperature of the dense gas to a level at which the protostellar accretion disc becomes stable.
Over time, dark matter annihilation will destroy all of the H$_2$ within the vicinity of the central protostar, making further fragmentation impossible. At distances larger than around 1,000~AU, however, the effects of dark matter annihilation are less pronounced and fragmentation can occur in some cases, typically resulting in the formation of a wide binary
system.
\end{enumerate}

\section*{Acknowledgements}
We gratefully acknowledge Thomas Greif for providing the initial conditions used for this work. R.J.S, DRGS and R.S.K.\ acknowledge support from the DFG via the SPP 1573 {\em Physics of the ISM} (grants SM321/1-1, KL 1358/14-1 \& SCHL 1964/1-1). R.J.S. and R.S.K. \ also acknowledges the support of the Landesstiftung Baden-W\"urttemberg under research contract P-LS-SPll/18 ({\em Internationale Spitzenforschung ll}) and the SFB 881 {\em The Milky Way System} subprojects B1, B2 and B4. DRGS thanks for funding via the SFB 963/1 ''Astrophysical Flow Instabilities and Turbulence".

\bibliography{Bibliography}

\end{document}